\documentclass[12pt]{iopart}
\usepackage{graphicx}
\usepackage{hyperref}
\usepackage{color}
\begin{document}
\title[Crystal field coefficients of Y analogues of RE-TM magnets using PAW-DFT]
{Crystal field coefficients for yttrium analogues of rare-earth/transition-metal
magnets using density-functional theory in the projector-augmented wave formalism}
\author{Christopher E.\ Patrick and
Julie B. Staunton}
\address{Department of Physics, University of Warwick,
Coventry CV4 7AL, UK}
\ead{c.patrick.1@warwick.ac.uk}

\begin{abstract}
We present a method of calculating crystal field coefficients of
rare-earth/transition-metal (RE-TM) magnets within density-functional theory (DFT).
The principal idea of the method is to calculate the crystal field
potential of the yttrium analogue (``Y-analogue'') of the RE-TM
magnet, i.e.\ the material where the lanthanide elements have been
substituted with yttrium.
The advantage of dealing with Y-analogues is that the methodological
and conceptual difficulties associated with treating the highly-localized
4$f$ electrons in DFT are avoided, whilst the nominal valence electronic
structure principally responsible for the crystal field is preserved.
In order to correctly describe the crystal field potential in the core
region of the atoms we use the projector-augmented wave formalism of
DFT, which allows the reconstruction of the full charge density and electrostatic
potential.
The Y-analogue crystal field potentials are combined with radial 4$f$ charge
densities obtained in self-interaction-corrected calculations on the lanthanides
to obtain crystal field coefficients.
We demonstrate our method on
a test set of 10 materials comprising
9 RE-TM magnets and elemental Tb.
We show that the calculated easy directions of magnetization agree
with experimental observations, including a correct description
of the anisotropy within the basal plane of Tb and NdCo$_5$.
We further show that the Y-analogue calculations
generally agree quantitatively with previous calculations using
the open-core approximation to treat the 4$f$ electrons,
and argue that our simple approach may be useful for large-scale
computational screening of new magnetic materials.
\end{abstract}
\submitto{\JPCM}
\maketitle

\section{Introduction}
Rare-earth/transition-metal (RE-TM) compounds, particularly
those containing neodymium, samarium and dysprosium, are the 
highest performing permanent magnets on the commercial 
market~\cite{Hirosawa2018,Gutfleisch2011}.
The key factor underpinning the success of these materials
is the possibility of obtaining a huge magnetocrystalline
anisotropy (MCA), i.e.\ a preferential direction for an object
to be magnetized independent of its macroscopic shape~\cite{Coey2011}, which
originates from the highly-localized 4$f$ electrons
of the lanthanide elements.
Specifically, the unfilled shell of 4$f$ electrons forms a
non-spherically symmetric charge cloud which sits in the 
(also non-spherically symmetric) crystal potential.
The interaction between the 4$f$ cloud and the crystal
potential, and the interaction between the spin and orbital
degrees of freedom of the 4$f$ electrons themselves, results in
a strong coupling of
the RE magnetism to the crystal potential~\cite{Gignoux1995}.
The spin-spin RE-TM interaction further
couples the magnetic moments of RE to those of the 
transition metals iron or cobalt~\cite{Brooks1989}.
The TM provides a large saturation
magnetization and high Curie temperature, which combine with the
high MCA to form an excellent permanent magnet~\cite{Buschow1977}.

Experimental research into RE-TM permanent magnets
has been carried out for over 50 years~\cite{Strnat1967,Strnat1972,
Kumar1988,Sagawa1984,Croat1984}.
Computational research, particularly that based on parameter-free, 
``first-principles'' methods, is a younger field by
comparison~\cite{Richter1998}.
However, the growth of computing power and a more widespread availability
of modelling codes has led to a rapid increase in recent years
of computational works aimed not only at understanding current 
RE-TM permanent magnets but also predicting the properties 
of new materials yet to be synthesized experimentally~\cite{Larson20032,Kashyap2003,Miura2014,
Matsumoto2014,Miyake2014,Harashima20152,Korner2016,Delange2017,Chouhan2017,
Patrick2017, Toga2018,Sakurai2018, Miyake2018,Tatetsu2018}.
Here a first-principles approach is highly desirable, since
the novel materials may require visiting a previously
unexplored parameter space where the reliability of 
empirical models is unknown.

However, an enduring challenge presented by RE-TM magnets 
for first-principles calculations is how to simultaneously describe
accurately the itinerant electrons of the TM and the highly-localized
4$f$ electrons of the lanthanide.
Practical implementations of density-functional theory (DFT)~\cite{Kohn1965}, 
a first-principles methodology which is highly popular due to its 
versatility and accuracy, require approximating the exchange-correlation
(XC) contribution to the total energy of the electrons.
Approximations based on the local spin density and
the homogeneous electron gas (LSDA/GGA) work very well for itinerant
electrons and form the basis of all widely-available DFT codes~\cite{Vosko1980,
Perdew1996}.
Unfortunately, LSDA/GGA XC functionals do not describe the lanthanide elements
well, with the 4$f$ electrons being too delocalized~\cite{Richter1991}.

In a previous publication~\cite{Patrick20182}, we discussed some different approaches which
attempt to correct the LSDA/GGA description of $4f$ electrons.
These approaches include the ``open core'' scheme~\cite{Brooks19912}, which effectively
removes the 4$f$ states from the valence band and freezes them in the
core, dynamical mean-field theory (DMFT)~\cite{Kotliar2006}, the self-interaction correction~\cite{Perdew1981},
and the LSDA/GGA+$U$ scheme~\cite{Ansimov1997}.
In that publication we chose the (local) self-interaction correction
(LSIC)~\cite{Lueders2005} and combined it with the 
disordered local moment (DLM) picture of finite
temperature magnetism~\cite{Gyorffy1985} to calculate the magnetization and Curie temperatures
of the entire series of RE-TM magnets with formula RECo$_5$~\cite{Patrick20182}.
However, using the same approach to calculate the MCA is problematic,
because current LSIC and DLM implementations employ a spherical
approximation for the potential at the RE site.
The most important contribution to the MCA, i.e.\ from the crystal potential,
is therefore incorrectly described~\cite{Hummler1996}.

If we wish to continue with our LSIC/DLM approach to obtain
a comprehensive picture of RE-TM magnets, it is apparent that we
must augment the original calculations to account for 
the non-spherical potential at the RE site.
Fortunately, an entire theoretical framework has been developed 
to describe the effects of this asphericity, namely
crystal field (CF) theory~\cite{Bleaney1953,Griffithbook,Newman1989,Kuzmin2008}.
Here, the potential at the RE site is expanded in terms
of angular functions, and matrix elements with radial 4$f$
wavefunctions are quantified in terms of CF coefficients.
Knowledge of the CF coefficients, combined with a description
of the finite temperature TM magnetism, can lead to a very
detailed picture of RE-TM magnetism~\cite{Kuzmin2008}.

The CF coefficients can be regarded as empirical parameters
used to fit experimental data~\cite{Tiesong1991}, but they can also be calculated.
The seminal crystal field models were based on the potential set up 
by arrays of point charges~\cite{Bleaney1953,Griffithbook}.
The advent of DFT allowed the CF coefficients to be calculated
from first principles~\cite{Richter1998}, originally from the electrostatic potential set
up by the charge density~\cite{Richter1995}, or later by also including the
XC contribution~\cite{Miyake2014,Novak1996,Steinbeck1996,Harashima2015}.
As usual, the 4$f$ electrons require special treatment, most frequently
via the open core scheme~\cite{Richter1998}.
Recent work has also seen the development of sophisticated techniques 
using DMFT to evaluate CF coefficients~\cite{Delange2017} or 
formulating CF theory in a Wannier basis~\cite{Novak2013}.

One reason that a number of different approaches can be found in the literature
is that CF theory is essentially empirical, and there is not a unique way of combining
it with non-empirical DFT.
In particular, in CF theory the 4$f$ electrons are spectators which
feel the crystal field but do not themselves influence it~\cite{Buck1997}.
But in a standard DFT calculation the CF potential contains contributions
from all electrons, meaning that special treatment of the 4$f$ electron
density (e.g.\ removing the non-spherical components~\cite{Miyake2014}) is required.

Here, we go a step further and present a method where we
\emph{completely} remove the 4$f$ electrons
from the crystal field,
by calculating the CF potential
of the yttrium (Y)-analogue of the RE-TM magnet.
Here ``Y-analogue''
means that have we have replaced all lanthanide atoms with yttrium;
for instance, the Y-analogue of Sm$_2$Co$_{17}$ is Y$_2$Co$_{17}$.
This approach is free from the complications of treating the XC energy of the 4$f$ electrons,
and produces a potential which does not contain 4$f$ contributions.

Of course, the validity of the approach depends on whether the valence electrons 
of the lanthanide are well represented by Y.
As a bare minimum, the lanthanide must be in a 3+ state, so that the nominal
valence configurations agree.
For the widely used RE-TM magnets we believe this requirement to be well satisfied~\cite{Richter1998},
but some care will be required in novel materials if the lanthanide is thought
to undergo valence fluctuations~\cite{Miyake2018}
(for such materials,
the open-core approximation would encounter the same problem as the Y-analogue model).
In addition, our calculations on the RE$^{3+}$Co$_5$ compounds revealed a
small contribution from $f$-type states around the Fermi level~\cite{Patrick20182} which will be 
missing for the Y-analogue, as will be any effects due to the (spherically-symmetric) spin polarization
of the 4$f$ electrons.
Therefore the use of the Y-analogue must be considered
an approximation.
However, the advantage of making this approximation is then we are free to 
compute the potential using common LSDA/GGA functionals, 
using widely available DFT codes.

Our manuscript describes the implementation of the Y-analogue scheme, focusing particularly
on two points.
The first is that if the chosen DFT code is not an ``all-electron'' code, i.e.\ 
it does not treat core and valence electrons on the same footing, the charge
density and potential will be incomplete in the core region
of the atoms.
By performing the calculations within the projector-augmented wave (PAW) formalism~\cite{Blochl1994},
as found in a number of popular codes including Quantum Espresso~\cite{Giannozzi2009}, 
\texttt{GPAW}~\cite{Enkovaara2010}
and the Vienna Ab initio Simulation Package (VASP)~\cite{Kresse1999}, it is possible to restore
this core contribution as a post-processing step.
In Secs.~\ref{sec.paw}--\ref{sec.pawdens} we describe how this is done.

The second point is that in order to calculate CF coefficients, we
require the radial distribution of the 4$f$ electrons.
As we discuss in Section~\ref{sec.radial}, here we use the spherically symmetric
4$f$ charge density obtained from a separate LSIC calculation.
Therefore our method consists of two separate strands, calculated with different
codes: one code provides the potential $V({\bf r})$ for the Y-analogue and the other 
code provides the radial density $n^0_{4f}(r)$ for the spherically-symmetric approximated
RE-TM compound.
Multiplying the two quantities and integrating yields the CF coefficients.

We have used our method to calculate CF coefficients for 10 materials, comprising 
9 RE-TM magnets and elemental Tb.
Where data is available we compare our calculations to previous work.
The data demonstrate all of the correct qualitative trends with respect to experiment,
and are in good quantitative agreement with open-core calculations.
We therefore present the method as an approximate but simple scheme 
to calculate CF coefficients.

The rest of our manuscript is organized as follows.
In Section~\ref{sec.theory} we start with a general overview of the crystal field
picture and then discuss the specifics of obtaining the potential in the PAW
formalism.
We also discuss the calculation of the spherically-symmetric 4$f$ density
and some of the conventions regarding CF coefficients.
In Section~\ref{sec.results} we present the calculated CF coefficients and 
compare to data previously published.
Finally in Section~\ref{sec.conclusions} we present our conclusions
and discuss potential future developments.

\section{Theory}
\label{sec.theory}
\subsection{Crystal field picture}
As comprehensively explained in Ref.~\cite{Kuzmin2008}  (and references
therein), crystal field theory describes
atomic-like electrons, which are eigenstates of a central potential
and characterized by a set of quantum numbers $|LSJM_J\rangle$,
perturbed by the CF potential $V({\bf r})$.
In the simplest case of a single magnetic sublattice subject to an external
field ${\bf B}$ with a spin-orbit coupling quantified by $\lambda$, 
the CF Hamiltonian is~\cite{Kuzmin2008}
\begin{equation}
\label{eq.ham}
\hat{H} = \lambda \hat{\bf L}\cdot \hat{\bf S} +  \mu_B (\hat{\bf L} + 2\hat{\bf S}) \cdot {\bf B} + \sum_iV({\bf r_i}).
\end{equation}
Here ${\bf r_i}$ denotes the position of a 4$f$ electron.
$V({\bf r})$ can be conveniently expanded in terms
of angular functions centred on the RE site.
Using the (complex) spherical harmonics $Y_{lm}({\bf \hat{r}})$,
this expansion is
\begin{equation}
V({\bf r}) = \sum_{lm}  V_{lm}(r) Y_{lm}({\bf \hat{r}}).
\label{eq.expansion}
\end{equation}
Matrix elements of $\sum_iV({\bf r_i})$ are 
products of radial and angular parts.
The angular parts are rewritten and evaluated in terms of operators, and
result in the appearance of Stevens coefficients $\alpha_J$, $\beta_J$
and $\gamma_J$, for $l$ = 2, 4, 6 respectively~\cite{Stevens1952}.
The radial part of the matrix element forms the CF coefficient,
which is the quantity that we aim to calculate in this work:
\begin{equation}
B_{lm} =  \left(\frac{2l+1}{4\pi}\right)^{\frac{1}{2}}\int r^2  n^0_{4f}(r) V_{lm}(r) dr
\label{eq.Blm}
\end{equation}
The sign has been defined such that a negative $V_{lm}(r)$ is attractive to 
an electron, which is the opposite of a conventional electrostatic potential~\cite{Kuzmin2008}.
$n^0_{4f}(r)$ is a spherically-symmetric charge density associated
with $4f$ electrons which we discuss in 
Section~\ref{sec.radial},
while we discuss the $(2l+1)/(4\pi)$ prefactor in Section~\ref{sec.BandA}.
First, we focus on calculating $V_{lm}(r)$.

\subsection{Kohn-Sham potential}
As mentioned in the Introduction, the original formulation
of CF theory supposed the perturbing potential
$V({\bf r})$ to be electrostatic in origin~\cite{Bleaney1953}.
However, in DFT the many-body system of interacting electrons
is mapped onto non-interacting Kohn-Sham (KS) electrons, which
experience both an electrostatic and an exchange-correlation
potential.
As discussed in Ref.~\cite{Richter1998},
it is reasonable to include
the XC contribution to the CF potential.
However, since the XC potential is spin dependent
one must either average over spins~\cite{Richter1998} or introduce
spin-dependent CF coefficients~\cite{Delange2017}.
To keep things general, we take the latter option and slightly
modify (\ref{eq.Blm}) so that there is a dependence on 
spin $\sigma (= \uparrow,\downarrow)$:
\begin{equation}
B^\sigma_{lm} =  \left(\frac{2l+1}{4\pi}\right)^{\frac{1}{2}}\int r^2  n^0_{4f}(r) V^\sigma_{lm}(r) dr
\label{eq.Bspin}
\end{equation}
where, inverting (\ref{eq.expansion}) and inserting the KS potential,
\begin{eqnarray}
V_{lm}^\sigma(r) &=&  \int V^\sigma_\mathrm{KS}({\bf r})  Y^*_{lm}({\bf \hat{r}})  d{\bf \hat{r}} \nonumber \\
                 &=&  \int [V_\mathrm{H}({\bf r}) + V^\sigma_\mathrm{XC}({\bf r})] Y^*_{lm}({\bf \hat{r}})  d{\bf \hat{r}}.
\label{eq.Vlm}
\end{eqnarray}
In our approach $V^\sigma_\mathrm{KS}({\bf r})$ is the
self-consistent KS potential of the Y-analogue of the RE-TM magnet.
Above we have split the KS potential into the electrostatic
potential $V_\mathrm{H}({\bf r})$ (which includes electrostatic electron-electron
and electron-nuclear
interactions),
and the XC potential $V^\sigma_\mathrm{XC}({\bf r})$.

\subsection{PAW}
\label{sec.paw}
In principle, if one performs an all-electron DFT calculation (particularly
using atom-centred basis sets)
then extracting $V_{lm}^\sigma(r)$ should be straightforward.
However, calculations which make some distinction between core
and valence electrons, e.g.\ those using pseudopotentials, do not deal directly
with $V^\sigma_\mathrm{KS}({\bf r})$ but rather with ``pseudized'' potentials
and densities.
The advantage of such schemes is that they remove the large computational
effort required to describe the rapidly varying electronic wavefunctions
in the core region~\cite{Martinbook}.

The projector-augmented wave (PAW) formalism is a popular method to perform
such calculations~\cite{Blochl1994}.
The full electronic+nuclear charge density is replaced with a 
pseudo-density $\tilde\rho({\bf r})$, which has an associated pseudo-electrostatic
potential which solves the Poisson equation 
$\nabla^2\tilde{V}_\mathrm{H}({\bf r}) = -4\pi\tilde\rho({\bf r}) $ (Hartree atomic units).
The difference between the full and pseudo-density is denoted $\Delta \rho({\bf r})$.
For each atom, an augmentation region is defined, which is an atom-centred sphere
with a radius of approximately 1--2.5 Bohr radii depending on the atom~\cite{GPAWsite}.
Crucially, outside the augmentation spheres the full and pseudo-densities are identical
(i.e.\ $\Delta \rho({\bf r})$ vanishes).
Inside, the pseudo-density is constructed such that it has the same multipole moments
as the full density.
Therefore, the full and pseudo-electrostatic potentials also match each other outside
the augmentation spheres~\cite{Blochl1994,Enkovaara2010}.

We thus write down a PAW version of (\ref{eq.Vlm}):
\begin{eqnarray}
V_{lm}^\sigma(r) &=& \int [\tilde{V}_\mathrm{H}({\bf r}) + 
V^\sigma_\mathrm{XC}(n^\uparrow({\bf r}),n^\downarrow({\bf r}))] Y^*_{lm}({\bf \hat{r}}) d{\bf \hat{r}} \nonumber \\
&&+ \Delta V_{\mathrm{H}lm}(r).
\label{eq.paw}
\end{eqnarray}
$\Delta V_{\mathrm{H}lm}(r)$ is the angular-resolved correction to the
electrostatic potential.
Note that we have now also specialized to the case that the XC potential
depends only on the spin densities at the point ${\bf r}$, i.e.\ the LSDA~\cite{Vosko1980}.
The all-electron spin-density is given by
\begin{equation}
n^\sigma({\bf r}) = \tilde{n}^\sigma({\bf r}) + \sum_{lm} \Delta n^\sigma_{lm}(r) Y_{lm}({\bf r}).
\label{eq.density}
\end{equation}
where  $\tilde{n}^\sigma({\bf r})$ is the electronic, spin-resolved contribution 
to the pseudo-density and $\Delta n^\sigma_{lm}(r)$ are the atom-centred corrections.

\subsection{Angular expansions of pseudized quantities}

To obtain the angular expansions of the pseudo-electrostatic potential  $\tilde{V}_{\mathrm{H}lm}(r)$ 
and the electronic pseudo-density $\tilde{n}_{lm}^\sigma(r)$, 
we use a Lebedev grid~\cite{Lebedev1999} of 5810 points on the unit sphere with vectors
${\bf \hat{r}_i}$ and weights $w_i$.
For example,
\begin{equation}
\tilde{V}_{\mathrm{H}lm}(r) 
= 4 \pi \sum_i w_i \tilde{V}_\mathrm{H}({\bf \hat{r}_i} r) Y^*_{lm}({\bf \hat{r}_i}).
\end{equation}
PAW codes generally include a postprocessing option to output $\tilde{V}_\mathrm{H}({\bf r})$
or $\tilde{n}^\sigma({\bf r})$ on a grid, and the smoothness of the data allows one to obtain
their values at arbitrary ${\bf r}$ through interpolation.

\subsection{Correction to the pseudo-electrostatic potential}

Recalling that $\Delta \rho({\bf r})$ vanishes outside the augmentation sphere, 
the correction to the electrostatic potential is given by
\begin{equation}
\Delta V_{\mathrm{H}}({\bf r}) = \int d{\bf r'} \frac{\Delta \rho({\bf r'})}{|{\bf r} - {\bf r'}|}.
\end{equation}
The angular expansion of $\Delta V_{\mathrm{H}}({\bf r})$ is therefore
\begin{equation}
\Delta V_{\mathrm{H}lm}(r) = 
\left( \frac{4\pi}{2l+1}\right)
 \int r'^2 dr' \Delta\rho_{lm}(r')\frac{r^l_<}{r^{l+1}_>},
\end{equation}
with $r_<$ and $r_>$ respectively denoting the lesser and greater
of $r'$ and $r$.

\subsection{Correction to the pseudo-density $\Delta\rho_{lm}(r)$}
\label{sec.pawdens}
$\Delta\rho({\bf r})$ consists of two contributions~\cite{Enkovaara2010}.
The first is from the nuclei, and requires replacing the soft
``compensation charges'' (which ensure the multipole moments
of the full and pseudo-density agree~\cite{Blochl1994}) with the point charge
at the origin with a $-Z/r$ potential.
The second contribution is the correction to the electron
pseudo-density $\Delta n({\bf r})$, which restores the
rapid variation of the electron density close to the origin
and also replaces the soft pseudo-core density with the
contribution from the atomic-like core states.
The resulting expression involves a number of PAW quantities 
and we give it in \ref{app.density}.
Here, we simply stress that all the quantities required are either
already included in the PAW datasets or computed during the SCF calculation,
so the computational effort required to obtain $\Delta\rho({\bf r})$ is minimal.

As well as using $\Delta\rho({\bf r})$ to calculate $\Delta V_{\mathrm{H}}({\bf r})$,
the spin-resolved all-electron density  (\ref{eq.density})
allows the computation of the XC potential in the same
angular representation.
Thus, using the steps described above, \ref{app.density} and (\ref{eq.paw}),
the angular resolved CF potential $V_{lm}^\sigma(r)$ 
appearing in (\ref{eq.Bspin}) is obtained.

\subsection{The spherically-symmetric 4$f$ charge density}
\label{sec.radial}
We now consider the other ingredient required to calculate the CF
coefficients, which is the spherically-symmetric 4$f$ charge density $n^0_{4f}(r)$.
In CF theory
$n^0_{4f}(r)$ originates from
single-particle eigenfunction of the unperturbed central potential which enters
the matrix element of the CF potential, $n^0_{4f}(r) = |\psi_{4f}(r)|^2$~\cite{Kuzmin2008},
which is normalized as
\begin{equation}
\int r^2 n^0_{4f}(r) dr = 1.
\label{eq.normalize}
\end{equation}

Of course, the Y-analogue model used to obtain  $V_{lm}^\sigma(r)$
does not provide $n^0_{4f}(r)$, since the stated aim of the model
was to remove the 4$f$ electrons.
However, an approach which aligns closely to the CF picture
is to perform an atomic-like calculation for a spherically-symmetric
potential, where the potential corresponds to the RE atom
embedded in the crystal.
Here we describe such an approach, based on scattering theory
and the local self-interaction correction (LSIC)~\cite{Lueders2005}.

The LSIC is an implementation of the SIC within the multiple-scattering,
Korringa-Kohn-Rostoker (KKR) Green's function formalism of DFT~\cite{Gyorffy1979}.
In particular, the KS potential is by construction in ``muffin tin'' form,
i.e.\ the potential is spherically symmetric within non-overlapping atom-centred
spheres and surrounded by a flat potential interstitial region (it is also possible
to use the ``atomic sphere'' construction, which removes the interstitial
region and allows the overlap of spheres)~\cite{Gyorffy1979}.
Previously we used such calculations (which are done at the scalar-relativistic level)
as starting points for fully-relativistic DLM calculations on RE-TM compounds for
the entire RECo$_5$ series from Y--Lu inclusive~\cite{Patrick20182}.

The scalar-relativistic potential obtained for the RE site from a self-consistent 
LSIC calculation can be inserted into the problem of the scattering of a free
electron by an isolated, spherically-symmetric and finite-ranged potential.
Apart from the regular and irregular solutions $Z$ and $J$, the central 
quantity in such a problem is the $t$-matrix~\cite{Ebert2011}.
The Green's function of the scattered electron is
\begin{eqnarray}
G({\bf r},{\bf r'},E) &=& Z({\bf r},E) t(E)Z^\times({\bf r'},E) \nonumber \\
&&- Z({\bf r},E)J^\times({\bf r'},E)
\label{eq.GF}
\end{eqnarray}
where $\times$ denotes a left-hand solution to the radial equation,
and matrix multiplication over angular indices has been implied (see
Refs.~\cite{Ebert2011,Tamura1992} for details).
The single-particle density $n_\mathrm{SP}$ is obtained from the Green's function as
\begin{equation}
n_\mathrm{SP}({\bf r}) = -\frac{1}{\pi}   \mathrm{Tr}  \int_C G({\bf r},{\bf r},E)dE  
\label{eq.GFn}
\end{equation}
The contour $C$ is a rectangle in the complex plane which encloses the energies 
of the bound 4$f$ states, which
in an LSIC calculation typically sit $\sim$1~Ry below the Fermi level~\cite{Patrick20182}.
For the light lanthanides (i.e.\ atomic numbers smaller than Gd) all of
the bound $4f$ states are included.
For the heavy lanthanides only the states belonging to the unfilled 
spin subshell are enclosed in the contour, since these states are the
ones responsible for the crystal field effects.

Since $Z$ and $J$ have the form of radial functions multiplied by spherical
harmonics, it is quite straightforward to use (\ref{eq.GF}) and (\ref{eq.GFn}) 
to extract the spherically-symmetric part of  $n_\mathrm{SP}$.
We then normalize this function to satisfy (\ref{eq.normalize}) and
set the result equal to $n^0_{4f}(r)$, i.e.\
\begin{equation}
n^0_{4f}(r) = \frac{\int n_\mathrm{SP}({\bf r}) d{\bf \hat{r}}}{ \int_{\Omega_\mathrm{MT}} n_\mathrm{SP}({\bf r'}) d{\bf r'}}.
\label{eq.4fdens}
\end{equation}
The normalization in (\ref{eq.4fdens}) has been
performed within the muffin tin sphere $\Omega_\mathrm{MT}$.
Accordingly, when calculating the CF coefficients from (\ref{eq.Bspin})
the integral is also performed up to the muffin tin radius.

\begin{figure}
\centering
\includegraphics{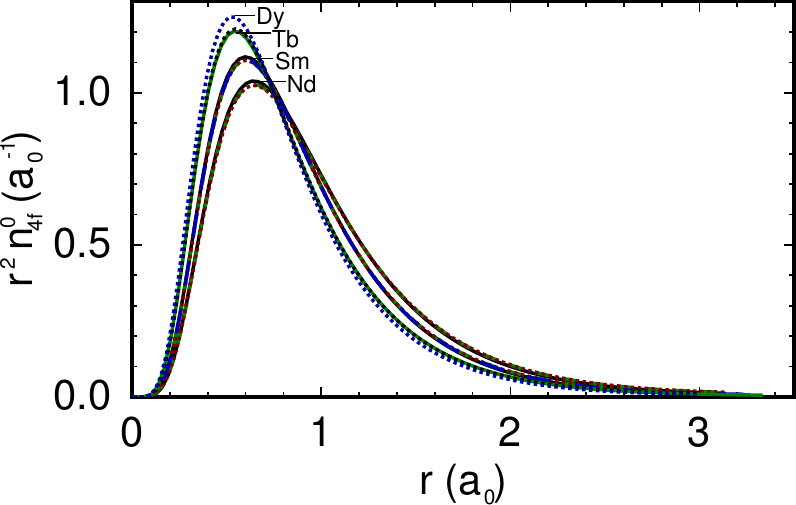}
\caption{
The spherically-symmetric 4$f$ electron density 
$n^0_{4f}(r)$ calculated for 10 compounds:
NdCo$_5$, SmCo$_5$, Tb (solid lines);
NdFe$_{12}$, SmFe$_{12}$ (dashed);
NdFe$_{12}$N, SmFe$_{12}$N, TbFe$_2$, DyFe$_2$ (dotted);
Sm$_2$Co$_{17}$ (long dash).
The data fall onto four distinct curves depending
on the lanthanide.
\label{fig.felec}
}
\end{figure}
In Fig.~\ref{fig.felec} we plot  $n^0_{4f}(r)$ obtained for
each of the 10 materials in our test set (more details regarding the test
set can be found in Sec.~\ref{sec.resultsmaterials}).
The most notable feature of Fig.~\ref{fig.felec} is that $n^0_{4f}(r)$
depends on the lanthanide but not on the host compound.
As a result, the 10 curves are effectively reduced to 4, corresponding
to the number of lanthanides in the test set (Nd, Sm, Tb and Dy).

Physically, the observation that $n^0_{4f}(r)$ depends on the lanthanide
but not the host compound is consistent with the picture of
atomic-like 4$f$ electrons.
Practically, this behaviour could be useful for studies screening a
large number of candidate compounds based on their CF coefficients.
To a good approximation, Fig.~\ref{fig.felec} supports the idea
that one need not recalculate  $n^0_{4f}(r)$ for every compound,
but rather predefine one function for each lanthanide to be used
for the full set of candidates.
Then the computational work would be restricted to calculating only the 
potential $V_{lm}^\sigma(r)$ for the Y-analogues.
For the current manuscript, however, we have recalculated $n^0_{4f}(r)$
for each compound.

\subsection{``$B$'' and ``$A$'' CF coefficients and axis orientation}
\label{sec.BandA}

Since we have chosen to expand the potential in terms of complex spherical
harmonics (\ref{eq.expansion}), it is natural to work with the CF
coefficients conventionally labelled $B$.
These coefficients correspond to expanding the potential with Wybourne
operators, which are related to the spherical harmonics by the 
prefactor appearing in (\ref{eq.Blm})~\cite{Newman1989,Kuzmin2008}.
Within this normalization one can distinguish ``real'' and ``imaginary''
CF coefficients depending on the relationship between $B_{lm}$ and
$B_{l-m}$.
For the materials considered here we have chosen the crystal axes such
that the imaginary CF coefficients are zero, so $B_{lm}$ = $B_{l-m}$.

In the past and present CF literature it is more common to find CF
coefficients $[A_{lm}\langle r^l\rangle]$ which are based on Stevens operators~\cite{Newman1989}.
Ref.~\cite{Newman1989} gives a table of factors $\alpha_{lm}$ which allow the conversion
$[A_{lm}\langle r^l\rangle] = \alpha_{lm} B_{lm}$ to be performed.
To facilitate comparison with previous work we perform this conversion and 
report our results using the Stevens form.
We stress that despite the notation, $[A_{lm}\langle r^l\rangle]$ should
be considered a single quantity in our calculations (hence the square brackets) and cannot be
be factored into $A_{lm}$ and  $\langle r^l\rangle$.

We also point out an issue relevant to materials with non-orthogonal axes,
e.g.\ hexagonal crystals.
In the expansion of (\ref{eq.expansion}), the conventional relation
between polar and cartesian co-ordinates applies, i.e.\ ${\bf \hat{r}} = 
(\sin\theta\cos\phi,\sin\theta\sin\phi,\cos\theta)$.
For cubic and tetragonal crystal structures, it makes sense to choose
the crystal axes to coincide with the three cartesian directions.
However, for a hexagonal system with the $c$ axis pointing in the $z$
direction, one can choose a crystal axis also to point in the 
$x$ or $y$ direction.

\begin{figure}
\centering
\includegraphics{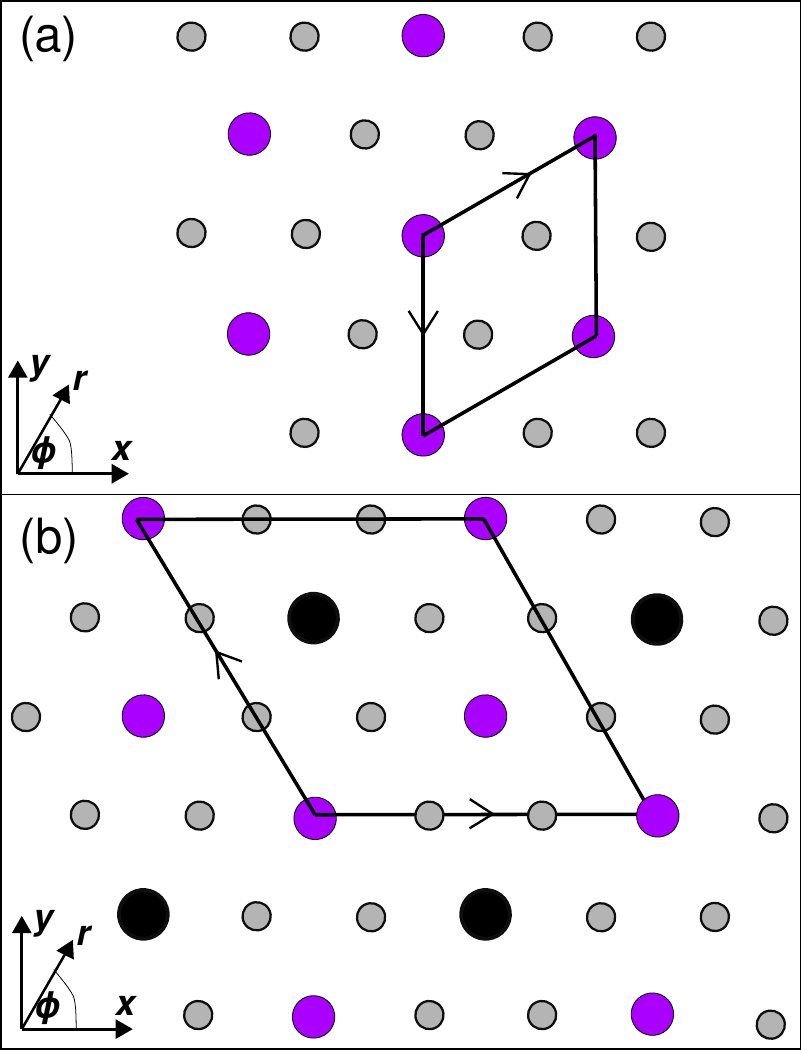}
\caption{
(a) The RECo$_5$ (CaCu$_5$) crystal structure~\cite{Kumar1988}, 
showing just the $ab$ plane containing the RE atom.
The hexagonal unit cell according to the convention of Ref.~\cite{BradleyCracknell}
is shown with the cartesian axes and polar co-ordinate representation.
(b) The RE$_2$Co$_{17}$ (Th$_2$Zn$_{17}$) crystal structure.
Notice how the unit cell is rotated compared to (a), but the hexagonal array
of Co atoms surrounding the RE are in the same orientation.
Grey circles represent Co, purple RE and black represents pairs of Co
atoms (dumbbells).
\label{fig.hex}
}
\end{figure}

The reason that such a choice has an impact on CF coefficients,
specifically those with nonzero $m$, is illustrated in Fig.~\ref{fig.hex}(a).
Here the SmCo$_5$ structure is shown, focusing on the $ab$ plane
containing the Sm atoms.
Figure~\ref{fig.hex}(a) obeys the convention given in the book by Bradley and 
Cracknell~\cite{BradleyCracknell},
so that one of the hexagonal lattice vectors is parallel to the $-y$ direction.
Such a system will in general have a nonzero $B_{66}$, which
is also equal to $B_{6-6}$.
However, one could construct a hexagonal co-ordinate
system with a hexagonal vector parallel to the $x$ direction,
corresponding to a 90$^\circ$ rotation in the $ab$ plane.
Since $Y_{6\pm6}\propto e^{\pm6i\phi}$, $B_{66}$ and $B_{6-6}$
would flip sign with this choice of axes.
Physically, the difference occurs because in
one case a vector with $\phi=0$ and its origin at the RE site
points towards the nearest neighbour Co atom,  while
for the other case the vector points exactly between two.

In this work, we mainly follow the convention of Fig.~\ref{fig.hex}(a)
when dealing with hexagonal systems.
The key exception is Sm$_2$Co$_{17}$, where we have a hexagonal
vector pointing along the $x$ direction [Fig.~\ref{fig.hex}(b)].
This choice of axes for Sm$_2$Co$_{17}$ ensures that the imaginary $B$ coefficients
are zero, and actually gives a local environment for the Sm
atoms which is more similar to SmCo$_5$ [i.e.\ nearest in-plane
Co atoms lying close to the $\phi=0$ direction, which can be seen
by comparing Figs.~\ref{fig.hex}(a) and (b)].

We have drawn attention to this issue because 
not all codes agree on the cartesian orientation of hexagonal axes.
For instance, the \texttt{Hutsepot} KKR code~\cite{Daene2009} used to perform the LSIC
calculation uses  the orientation of Fig.~\ref{fig.hex}(a) for 
hexagonal crystals by default.
However, in the Quantum Espresso package~\cite{Giannozzi2009} 
or the Atomic Simulation Environment (ASE)~\cite{Larsen2017} hexagonal axes 
are defined with a lattice vector pointing parallel to the $x$ direction [Fig.~\ref{fig.hex}(b)].
With this in mind we think it useful when reporting crystal field
coefficients with nonzero $m$ for hexagonal structures to also
give the relationship between crystal axes and $\theta$ and $\phi$.

Finally, we note an additional potential source of confusion is that
in experimental literature there are $B$ coefficients
which are distinct from those introduced above.
These alternative $B$ coefficients are
the $[A_{lm}\langle r^l \rangle]$ quantities multiplied by the
appropriate Stevens coefficient $\alpha_J$, $\beta_J$ or $\gamma_J$,
depending on $l$ ~\cite{Radwanski1987}.
For clarity, we do not refer to these quantities any further here.

\subsection{Anisotropy constants from crystal field coefficients}
\label{sec.KfromA}

This work focuses on the calculation of CF coefficients.
However, experimental studies involving magnetization measurements
more commonly report anisotropy constants, which (for a uniaxial system) 
come from an expansion of the free energy
in terms of the magnetization direction ($\theta_0$, $\phi_0$):
\begin{eqnarray}
\label{eq.K}
E(\theta_0,\phi_0) &=& K_1 \sin^2 \theta_0 + K_2 \sin^4 \theta_0 + K_3 \sin^6 \theta_0 \nonumber \\
&&+ K_4 \sin^6\theta_0 \cos 6\phi_0.
\end{eqnarray}

For Tb, where there is no contribution from the TM, we can calculate the $K_i$
values by diagonalizing 
the Hamiltonian in~(\ref{eq.ham}) within the ground $J$ manifold
($L$=$S$=3, $J$ = 6).
We choose the magnitude of the external field in~(\ref{eq.ham})  to be very large (5000~T)
which constrains the magnetization to point along the chosen field direction ${\bf\hat B}$.
The zero temperature (ground state) energy is thus obtained as a function of magnetization angle, which can
be fitted to the expression~\ref{eq.K}.
In the case that the $l=2$ term dominates, the following relation is satisfied~\cite{Kuzmin2008}:
\begin{equation}
K_1 = -3 J (J-0.5) \alpha_J [A_{20}\langle r^2\rangle].
\label{eq.KA2}
\end{equation}

For a general RE-TM magnet it is not straightforward to relate CF coefficients to the $K_i$
constants.
This is because the CF coefficients only contain the contribution of the RE,
but the TM will also contribute to the anisotropy~\cite{Kuzmin2008}.
However, making the reasonable assumption that the RE provides the dominant
contribution, one can
use the CF coefficients to at least make qualitative statements about the
anisotropy.
For instance, multiplying the signs of the Stevens factor $\alpha_J$
and  $[A_{20}\langle r^2\rangle]$ gives the sign of negative $K_1$, assuming
the $l=2$ term is dominant (\ref{eq.KA2}).
Similarly, for cubic systems like the Laves phase compounds (where the first nonzero
CF coefficient has $l=4$), the product of $[A_{40}\langle r^4\rangle]$ and $\beta_J$ will
be positive for an easy [111] direction and negative for [001].
The latter behaviour can be verified by plotting the potential $B_{40}Y_{40}(\theta)
+ B_{4\pm4}Y_{4\pm4}(\theta)$ in the $\phi=0$ plane multiplied by the sign of the Stevens factor, 
and seeing if the result is maximal or minimal along the [001] direction.

\subsection{Computational details}

We calculated the potential  $V_{lm}^\sigma(r)$ including the PAW
corrections (\ref{eq.paw}) for the Y-analogues of RE-TM magnets
using the \texttt{GPAW} code, version 1.4.0~\cite{Enkovaara2010}.
We used the latest freely available \texttt{GPAW} PAW datasets (version 0.9.2)~\cite{GPAWsite}.
The calculations were performed using a plane wave basis set, expanding
the wavefunctions up to a maximum plane wave kinetic energy of 1200~eV.
The reciprocal space sampling was performed using $\Gamma$-centred $k$-point
meshes and converged for each material, and the Kohn-Sham states occupied according
to a Fermi-Dirac distribution with a width of 0.01~eV.
The exchange-correlation energy was modelled with the LSDA~\cite{Vosko1980}.
The PAW corrections (\ref{app.density}) were extracted  
using our own \texttt{Python} scripts.

To calculate the spherically-symmetric 4$f$ electron density 
$n^0_{4f}(r)$ we used the \texttt{Hutsepot} KKR code~\cite{Daene2009} to calculate
the spherically-symmetric potential at the RE site within the 
muffin-tin approximation.
The muffin tin radii are reported in Table~\ref{tab.structures}.
We used the LSDA + LSIC~\cite{Lueders2005} to model the XC energy, 
where the angular momentum channels were corrected according to the 
scheme described in Ref.~\cite{Patrick20182}.
Using this potential we solved the atomic problem on a logarithmic radial grid
to obtain the Green's function
and density [(\ref{eq.GFn}) and (\ref{eq.4fdens})].
Finally, $V_{lm}^\sigma(r)$ and  $n^0_{4f}(r)$ were combined to calculate
CF coefficients from (\ref{eq.Bspin}) and converted into 
$[A_{lm}\langle r^l\rangle]$ notation~\cite{Newman1989}.

\section{Results}
\label{sec.results}
\subsection{Materials considered}
\label{sec.resultsmaterials}

\begin{table*}
\caption{\label{tab.structures}  Structural parameters used in the calculations, with the Wyckoff
positions occupied by each inequivalent atom given.
Lattice parameters and muffin tin radii are in \AA, internal co-ordinates
of the indicated Wyckoff positions are dimensionless.}

\scriptsize
\begin{tabular}{@{}lllll}
\br
Material        &Space group;     &Lattice parameters;      &Internal co-ordinates                       &Source and reference\\
                &RE site symmetry &MT radius                &                                            &                    \\
\mr
SmCo$_5$        & 191 (P6/mmm)   &$a$ = 4.974, $c$ = 3.978    & Sm(1$a$), Co(2$c$), Co(3$g$)                  & Exp.\ at 5~K~\cite{AndreevHMM}\\
                & $D_{6h}$         & $r_\mathrm{MT}$=1.645    &                                               &                                    \\
NdCo$_5$        & 191 (P6/mmm)   &$a$ = 5.006, $c$ = 3.978    & Nd(1$a$), Co(2$c$), Co(3$g$)                  & Exp.\ at 5~K~\cite{AndreevHMM}\\
                & $D_{6h}$         & $r_\mathrm{MT}$=1.661    &                                               &                                    \\
Sm$_2$Co$_{17}$ & 166 (R$\bar{3}$m) &$a$ = 8.398, $c$ = 12.218& Sm(6$c$), Co(6$c$), Co(9$d$), Co(18$f$), Co(18$h$) & Exp.\ at 300~K~\cite{Herbst1982}\\
                & $C_{3v}$         & $r_\mathrm{MT}$=1.781    & $z_{6c,\mathrm{Sm}}$ = 0.343; $z_{6c,\mathrm{Co}}$ = 0.099;  $x_{18f}$ = 0.283;&(Nd$_2$Co$_{17}$; see text)\\ 
                &                  &                          & $(x,z)_{18h}$ = (0.171,0.486)                  &                     \\ 
NdFe$_{12}$     & 139 (I4/mmm)      &$a$ = 8.533, $c$ = 4.681 & Nd(2$a$), Fe(8$f$), Fe(8$i$), Fe(8$j$)         & DFT-GGA~\cite{Harashima2015}\\
                & $D_{4h}$         & $r_\mathrm{MT}$=1.667    & $x_{8i}$ = 0.359; $x_{8j}$ = 0.268             &                     \\ 
NdFe$_{12}$N    & 139 (I4/mmm)      &$a$ = 8.521, $c$ = 4.883 & Nd(2$a$), N(2$b$), Fe(8$f$), Fe(8$i$), Fe(8$j$)& DFT-GGA~\cite{Harashima2015}\\
                & $D_{4h}$         & $r_\mathrm{MT}$=1.667    & $x_{8i}$ = 0.361; $x_{8j}$ = 0.274             &                     \\ 
SmFe$_{12}$     & 139 (I4/mmm)      &$a$ = 8.497, $c$ = 4.687 & Sm(2$a$), Fe(8$f$), Fe(8$i$), Fe(8$j$)         & DFT-GGA~\cite{Harashima2015}\\
                & $D_{4h}$         & $r_\mathrm{MT}$=1.667    & $x_{8i}$ = 0.359; $x_{8j}$ = 0.270             &                     \\ 
SmFe$_{12}$N    & 139 (I4/mmm)      &$a$ = 8.517, $c$ = 4.844 & Sm(2$a$), N(2$b$), Fe(8$f$), Fe(8$i$), Fe(8$j$)& DFT-GGA~\cite{Harashima2015}\\
                & $D_{4h}$         & $r_\mathrm{MT}$=1.667                         & $x_{8i}$ = 0.361; $x_{8j}$ = 0.274             &                     \\ 
TbFe$_{2}$      & 227 (Fd$\bar{3}$m); $T_d$&$a$ = 7.341, $r_\mathrm{MT}$=1.589     & Tb(8$b$), Fe(16$c$)                            & Exp.\ at 300~K~\cite{AndreevHMM}\\
DyFe$_{2}$      & 227 (Fd$\bar{3}$m); $T_d$&$a$ = 7.338, $r_\mathrm{MT}$=1.589     & Dy(8$b$), Fe(16$c$)                            & Exp.\ at 300~K~\cite{AndreevHMM}\\
Tb              & 194 (P6$_3$/mmc) &$a$ = 3.606, $c$ = 5.698  & Tb(2$c$)                                       & Exp.\ at 300~K~\cite{Smidt1963}\\
                & $D_{3h}$         & $r_\mathrm{MT}$=1.764    &                                                &                     \\
\br
\end{tabular}
\end{table*}

In Table~\ref{tab.structures} we list the 10 materials considered in this work.
In our set of materials we have included the archetypal Sm-Co magnets SmCo$_5$~\cite{Strnat1967} 
and Sm$_2$Co$_{17}$~\cite{Strnat1972}, and
also examples of the 1:12 magnet class e.g.\ NdFe$_{12}$N, which are currently
the subject of research due to their potential as hard magnetic materials with reduced
RE content~\cite{Gabay2017}.
We have also included the Laves phase magnets TbFe$_2$ and DyFe$_2$ whose alloy is the
highly magnetostrictive Terfenol-D compound~\cite{Clark1973,Abbundi1977}, and also elemental Tb~\cite{Rhyne1967}.
Tb is an interesting test since the Y-analogue has no atoms in common with the original structure.

Table~\ref{tab.structures} also gives the structural parameters (lattice constants and internal co-ordinates) 
used in the calculations.
In order to be able to best compare our CF coefficients to previous calculations
we have where possible used the same structural parameters.
Accordingly, as indicated in Table~\ref{tab.structures} structural parameters have been
sourced both from experimental and computational works.
For Sm$_2$Co$_{17}$ we follow Ref.~\cite{Steinbeck1996} and use the structural parameters
of the related compound Nd$_2$Co$_{17}$~\cite{Herbst1982}, which avoids the experimental
difficulties associated with performing neutron experiments on Sm compounds.

As clearly explained in Ref.~\cite{Kuzmin2008}, depending on the point symmetry of
the RE site only certain CF coefficients will be nonzero.
Furthermore, for the purposes of calculating matrix elements for $f$ electrons we
need only calculate CF coefficients for even values of $l$, up to a maximum $l=6$.
In order to check that our computational method is robust we calculated the
decomposition of the potential (\ref{eq.paw}) for all $lm$ combinations and
confirmed that the only nonzero terms were those expected from the point symmetry~\cite{BradleyCracknell}.

\subsection{CF coefficients}
\begin{table*}
\caption{\label{tab.CFcoffs}  Calculated crystal field coefficients, in K (i.e.\ $[A_{lm}\langle r^l\rangle]/k_B$).  
The two numbers reported for each $lm$ combination corresponds to spin up/down respectively, and the bold
number corresponds to the spin of the asymmetric 4$f$ electron cloud, as discussed in the text.
}
\scriptsize
\begin{tabular}{@{}lllllllll}
\br
Material        &$[A_{20}\langle r^2\rangle]$   &$[A_{40}\langle r^4\rangle]$     
& $[A_{43}\langle r^4 \rangle]$    & $[A_{44}\langle r^4 \rangle]$     
& $[A_{60}\langle r^6 \rangle]$    
& $[A_{63}\langle r^6 \rangle]$   & $[A_{64}\langle r^6 \rangle]$   & $[A_{66}\langle r^6 \rangle]$      \\
\mr
SmCo$_5$        &-402/{\bf-400}  &-30/{\bf-22} &  ---        & ---          & 5/{\bf 4}  & ---        & ---        &-137/{\bf-115} \\
\ \ DMFT~\cite{Delange2017}&-313/{\bf-262}  &-40/{\bf-55} &--- & ---        & 35/{\bf 25}& ---        & ---        &-731/{\bf-593} \\
NdCo$_5$        &-421/{\bf-415}  &-36/{\bf-26} &  ---        & ---          & 6/{\bf 5}  & ---        & ---        &-174/{\bf-146} \\
Sm$_2$Co$_{17}$ &-199/{\bf-208}  &-20/{\bf-7}  &116/{\bf 108}& ---          & -1/{\bf-1} &-22/{\bf-25}& ---        &-54/{\bf -48}   \\      
\ \  open core~\cite{Steinbeck1996}&{\bf -194}&{\bf -15} &{\bf 74} & ---    & {\bf -2}   & {\bf -61}  & ---        &{\bf -139 }         \\
NdFe$_{12}$     &-116/{\bf-110}  &  4/{\bf 15} &  ---        &-205/{\bf-145}& 10/{\bf 7} & ---        &-29/{\bf-21}& ---           \\      
\ \  open core~\cite{Harashima2015}&{\bf -77}& ---   &  ---        & ---          & ---        & ---        & ---        & ---           \\
\ \ DMFT~\cite{Delange2017}&-71/{\bf-116} &-5/{\bf -1}& --- &-76/{\bf-270}  & 62/{\bf 54}& ---        &-224/{\bf-107}& --- \\      
NdFe$_{12}$N    & 188/{\bf 364}  & -62/{\bf-13}&  ---        &-161/{\bf-101}&-17/{\bf-16}& ---        &-1/{\bf-3}  & ---           \\      
\ \  open core~\cite{Harashima2015}&{\bf 367}& ---  &  ---        & ---          & ---        & ---        & ---        & ---           \\
\ \ DMFT~\cite{Delange2017}&477/{\bf 653} &75/{\bf 112}& --- &-105	/{\bf-141}  & 32/{\bf 63}& ---        &-65/{\bf-91}&  --- \\      
SmFe$_{12}$     &-100/{\bf -96}  &   1/{\bf 10}&  ---        &-154/{\bf-112}&  8/{\bf 6} & ---        &-21/{\bf-15}& ---           \\      
\ \  open core~\cite{Harashima2015}&{\bf -47} & ---  &  ---        & ---          & ---        & ---        & ---        & ---           \\
\ \ DMFT~\cite{Delange2017}&-184/{\bf-211} &-21/{\bf-18}& --- &-41	/{\bf-136}  & 45/{\bf 40}& ---        &-95/{\bf-58}&  --- \\      
SmFe$_{12}$N    & 272/{\bf 414}  & -47/{\bf-6} &  ---        &-121/{\bf-75} &-14/{\bf-13}& ---        &-2/{\bf-3}  & ---           \\      
\ \  open core~\cite{Harashima2015}&{\bf 371} & ---  &  ---        & ---          & ---        & ---        & ---        & ---           \\
\ \ DMFT~\cite{Delange2017}&195/{\bf 225} &78/{\bf 70}& --- &22/{\bf-91}    & 47/{\bf 25}& ---        &-97/{\bf-82}&  --- \\      
TbFe$_{2}$      &      ---       &  {\bf 28}/29&  ---        &{\bf 139}/143 & {\bf -2}/-2& ---        &{\bf 39}/38 & ---           \\      
DyFe$_{2}$      &      ---       &  {\bf 26}/26&  ---        &{\bf 128}/132 & {\bf -2}/-2& ---        &{\bf 34}/32 & ---           \\      
Tb              &-59/{\bf-60}    &  -4/\bf{-3} &  ---        & ---          & 4/{\bf 4}  & ---        &  ---       & 36/{\bf 36}   \\      
\br
\end{tabular}
\end{table*}

Table~\ref{tab.CFcoffs} gives the CF coefficients calculated for our set
of materials.
We also reproduce the values of CF coefficients calculated previously
in the literature for some of the materials~\cite{Delange2017,Steinbeck1996,Harashima2015}.

For our calculations, and Ref.~\cite{Delange2017},
two numbers are reported for each $lm$, corresponding
to the CF coefficient calculated using $V_{lm}^\uparrow(r)$ 
and $V_{lm}^\downarrow(r)$.
One of the two numbers has been written in bold, corresponding
to the coefficient that (in the zero temperature, scalar-relativistic
picture) is the potential felt by the partially filled 4$f$ spin subshell.
Expanding on this point, we first note that we have chosen the 
TM spins to point in the $\uparrow$ direction.
Furthermore, the RE-TM coupling is generally antiferromagnetic
between spins~\cite{Brooks1989}.
Therefore, for the light REs Nd and Sm the partially filled 4$f$ subshell
has $\downarrow$ spin character, so the relevant CF coefficient
is calculated using $V_{lm}^\downarrow(r)$.
For the heavy REs, the  4$f$ $\downarrow$ spin subshell is filled,
so $V_{lm}^\uparrow(r)$ is relevant for TbFe$_2$ and DyFe$_2$.
For elemental Tb we chose $\uparrow$ to
be the majority spin direction, so the unfilled 4$f$ spin
subshell is $\downarrow$.
We note that, except for NdFe$_{12}$N and SmFe$_{12}$N, we calculate the difference
between $V_{lm}^\uparrow(r)$  and $V_{lm}^\downarrow(r)$ to be rather small.
We now discuss each material in more detail.

\subsubsection{SmCo$_5$ and NdCo$_5$}

SmCo$_5$ crystallizes in the CaCu$_5$ structure, which is hexagonal with
one formula unit in the unit cell~\cite{Kumar1988}.
SmCo$_5$ is characterized by a large uniaxial anisotropy~\cite{Ermolenko1976},
which is understood in terms of CF theory based on the approximate relation (\ref{eq.KA2}) between
the first anisotropy constant and lowest order CF coefficient,
$K_1\propto -\alpha_J [A_{20}\langle r^2\rangle]$~\cite{Kuzmin2008}.
The positive value of 13/315 for the Stevens $\alpha_J$ coefficient of 
Sm$^{3+}$~\cite{Kuzmin2008} means that
SmCo$_5$ should have a large, negative value of  $[A_{20}\langle r^2\rangle]$~\cite{Buschow1974}.

Recently, Ref.~\cite{Delange2017} calculated the CF coefficients of SmCo$_5$ within
DMFT, and also provided a useful summary of past calculations and experimental measurements.
Experimentally measured values of  $[A_{20}\langle r^2\rangle]$ vary between -180 and -420~K,
whilst calculations found values between  -160 and -755~K.
There is no particular consensus about the other CF coefficients, although the computational
works agreed in finding $[A_{40}\langle r^4\rangle]$ to be negative and $[A_{60}\langle r^6\rangle]$
to be positive.
We reproduce the results of the DMFT calculations with our own results in Table~\ref{tab.CFcoffs}.

Our calculations based on the Y-analogue model of SmCo$_5$ 
find values of the $[A_{20}\langle r^2\rangle]$ coefficient (-402/-400~K) 
which fall into the previously reported experimental/calculated ranges.
The variation in sign of $[A_{20}\langle r^2\rangle]$, $[A_{40}\langle r^4\rangle]$ and $[A_{60}\langle r^6\rangle]$
also follows that observed in previous calculations.
Despite the different methodologies involved, our calculated values
are reasonably close to the DMFT results~\cite{Delange2017}, with the exception
of $[A_{66}\langle r^6\rangle]$.
This CF coefficient controls the basal plane anisotropy.
However, it does not affect the ground-$J$ multiplet of Sm~\cite{Kuzmin2008}.
Furthermore, higher order CF coefficients decay
more quickly with temperature~\cite{Kuzmin2008}.
Therefore, the effect of  $[A_{66}\langle r^6\rangle]$
on the strongly uniaxial SmCo$_5$ is expected to be difficult to observe,
and has not been measured experimentally~\cite{Buschow1974}.

For comparison we also considered the isostructural NdCo$_5$ compound.
Within the Y-analogue model, any differences between CF coefficients
calculated for SmCo$_5$ and NdCo$_5$ must be attributed to (a) the slightly
different lattice parameters (Table~\ref{tab.structures}) and (b) the different
4$f$ electron density (Fig.~\ref{fig.felec}).
We see from Table~\ref{tab.CFcoffs} that the calculated CF coefficients are actually
very similar for both materials.
However, Nd$^{3+}$ has a negative Stevens factor (-7/1089)~\cite{Kuzmin2008},
which means NdCo$_5$ should have an in-plane anisotropy (negative $K_1$),
This is exactly what is observed experimentally~\cite{Klein1975}.
Furthermore, unlike SmCo$_5$ the anisotropy within the basal plane has also
been measured experimentally, with the easy direction found to point between
Co atoms~\cite{Radwanski1987} [$\phi=30^\circ$ in Fig.~\ref{fig.hex}(a)].
This easy direction is indeed consistent with the negative $[A_{66}\langle r^6\rangle]$ which we calculate, 
since the Stevens $\gamma_J$ ($l$=6) coefficient of Nd$^{3+}$ is negative~\cite{Kuzmin2008}.

\subsubsection{Sm$_2$Co$_{17}$}

The magnet Sm$_2$Co$_{17}$ forms in the Th$_2$Zn$_{17}$ structure.
This structure is closely related to SmCo$_5$ by an ordered substitution of
one in three Sm atoms with a pair (dumbbell) of Co atoms~\cite{Kumar1988}.
Although there is a distortion of the local environment, the Sm atoms are still
surrounded by a hexagon of effectively coplanar Co atoms, as shown in Fig.~\ref{fig.hex}(b).
However, as is also clear from Fig.~\ref{fig.hex}(b) the dumbbells lower
the hexagonal symmetry so that CF coefficients with $m=\pm3$ are nonzero.
What is not shown in Fig.~\ref{fig.hex} is the tripling of the unit cell
along the $c$ direction compared to SmCo$_5$, due to the stacking of
two Sm atoms and a dumbbell.
Again we stress that our choice of axes in Fig.~\ref{fig.hex}(b) is unconventional
compared to Fig.~\ref{fig.hex}(a), but physically gives a closer correspondence
between the $(\theta,\phi)$ co-ordinate system and the positions of atoms in SmCo$_5$.

Now considering the calculated $[A_{lm}\langle r^l \rangle]$ in Table~\ref{tab.CFcoffs},
we see that introducing the Co dumbbells results in reduced coefficients
compared to SmCo$_5$.
In particular, $[A_{20}\langle r^2 \rangle]$ is halved.
The value of -208~K is still negative however, supporting uniaxial anisotropy
as is observed experimentally~\cite{Kumar1988}.
In terms of anisotropy within the basal plane,  $[A_{66}\langle r^6 \rangle]$
is negative like in SmCo$_5$, but is weaker by a factor of 2.

Ref.~\cite{Steinbeck1996} reported calculations of the CF coefficients of Sm$_2$Co$_{17}$,
modelling the Sm in the 3+ state with the open-core approximation.
We reproduce the calculated values in Table~\ref{tab.CFcoffs}, and find very
close agreement with our Y-analogue model.
The exception is for the $l=6$ coefficients, with Ref.~\cite{Steinbeck1996}
finding values a factor of 3 larger than our calculations.
As for SmCo$_5$, there is no experimental data for these coefficients.
However, we think that it is reasonable that our calculations show
a reduction in $[A_{66}\langle r^6 \rangle]$ for Sm$_2$Co$_{17}$ compared
to SmCo$_5$, due to the disrupted hexagonal symmetry.

\subsubsection{NdFe$_{12}$, NdFe$_{12}$N, SmFe$_{12}$, SmFe$_{12}$N}
Members of the 1:12 magnet class have a tetragonal, ThMn$_{12}$ crystal structure
with the RE surrounded by 3 TM sublattices.
The interstitial $2b$ positions can also be occupied by nonmetal atoms,
forming e.g.\ NdFe$_{12}$N~\cite{Gabay2017}.
The fourfold symmetry in the $ab$ plane gives rise to nonzero
CF coefficients with $m\pm4$.
However, it is the lowest order
coefficient $[A_{20}\langle r^2 \rangle]$ (which decays the slowest
with temperature) which is expected to have the largest effect
on the uniaxial anisotropy.
As for the materials above, a negative value of  $[A_{20}\langle r^2 \rangle]$
is expected to yield uniaxial anisotropy for Sm and planar anisotropy for
Nd, based on the sign of the Stevens coefficient.

Table~\ref{tab.CFcoffs} shows our calculated CF coefficients for
NdFe$_{12}$, NdFe$_{12}$N, SmFe$_{12}$ and SmFe$_{12}$N.
The most striking feature is the sign reversal of 
$[A_{20}\langle r^2 \rangle]$ upon nitrogenation, switching from
negative to positive.
These signs mean that SmFe$_{12}$ and NdFe$_{12}$N should have uniaxial anisotropy,
and NdFe$_{12}$ and SmFe$_{12}$N should be planar.
Nitrogenation also provides a negative contribution to 
$[A_{40}\langle r^4 \rangle]$, but weakens $[A_{44}\langle r^4 \rangle]$.
The $l=6$ CF coefficents are small for all of the materials.

Also in Table~\ref{tab.CFcoffs} we reproduce reported CF coefficients calculated
with DMFT~\cite{Delange2017} and with the open core approximation~\cite{Harashima2015}
(only the $l$=2 coefficients are reported in Ref.~\cite{Harashima2015}).
Focusing first on $[A_{20}\langle r^2 \rangle]$ we find that the three methods
calculate the same signs.
Furthermore there is close numerical agreement between our Y-analogue calculations
and the open core calculations~\cite{Harashima2015}, as for Sm$_2$Co$_{17}$.
However, there is no systematic level of agreement of these calculations with DMFT,
with for instance similar values calculated for NdFe$_{12}$ but variations up
to a factor of two for the other materials.
The agreement between the Y-analogue model and DMFT for the higher order
CF coefficients is similarly unsystematic, with for instance opposite
signs found for $[A_{40}\langle r^4 \rangle]$ but quite close agreement
for $[A_{44}\langle r^4 \rangle]$.
Like for SmCo$_5$, the DMFT calculations find larger values of $l$=6 coefficients
than in the Y-analogue model.

We note that both the DMFT and Y-analogue calculations find that nitrogenation
introduces a significant difference between the $\uparrow$ and $\downarrow$ CF coefficients.
At first sight, this observation is puzzling since N is nonmagnetic.
However, as shown in Ref.~\cite{Harashima2015}
the introduction of N strengthens the magnetization by approximately 2.5~$\mu_\mathrm{B}$
per formula unit, which our calculations find is due to an enhanced Fe magnetization
at the 8$f$ sites.
The 8$f$ sites sit halfway between RE planes, so it is not unreasonable that 
an enhanced spin polarization here will affect the $\uparrow$ and $\downarrow$ CF coefficients
by differing amounts.

Finally, comparing the Y-analogue calculations between Nd and Sm compounds we see that the
same qualitative behavior of the CF coefficients is observed.
However, unlike for SmCo$_5$ and NdCo$_5$ there are some numerical differences, 
particularly for the nitrided compounds where there is a large difference
in $c$ parameter for NdFe$_{12}$N and SmFe$_{12}$N.

\subsubsection{TbFe$_2$ and DyFe$_2$}
The Laves phase REFe$_2$ compounds are notable for their large room temperature
magnetostrictions, particularly the alloy Tb$_{0.27}$Dy$_{0.73}$Fe$_2$
(Terfenol-D)~\cite{Clark1973,Abbundi1977}.
Unlike the other materials in our test set the Laves phase (MgCu$_2$) structure is cubic,
yielding a zero $[A_{20}\langle r^2 \rangle]$ coefficient.
They are also key examples of RE-TM magnets
based on heavy REs with practical applications.

In Table~\ref{tab.CFcoffs} we give the CF coefficients
calculated using the Y-analogue model for TbFe$_2$ and DyFe$_2$.
We point out that although four coefficients are presented for
each material, only $[A_{40}\langle r^4 \rangle]$ and $[A_{60}\langle r^6 \rangle]$
are independent, with the cubic symmetry fixing the ratios $B_{44}:B_{40}$ and
$B_{64}:B_{60}$~\cite{BradleyCracknell}.
The first notable feature is that the calculated coefficients 
are almost identical for TbFe$_2$ and DyFe$_2$.
As a consequence, within CF theory any different behavior of the anisotropy 
must come from the Stevens factors.

The $\beta_J$ ($l$=4) factors are 2/16335 and -8/135135 for Tb$^{3+}$ and Dy$^{3+}$,
respectively.
The $\gamma_J$ coefficients also have opposite signs~\cite{Kuzmin2008}.
Therefore, one would expect TbFe$_2$ and DyFe$_2$ to have different
easy directions of magnetization, either [111] or [100].
Specifically, based on the positive  $l=4$ CF coefficients 
one would expect an easy direction of [111] for TbFe$_2$ and [100] for
DyFe$_2$ (Sec.~\ref{sec.KfromA}).
This behaviour is exactly what is observed experimentally; indeed
the easy directions of all of the heavy RFe$_2$ compounds follow
the sign of $\beta_J$~\cite{AndreevHMM}.

To our knowledge the CF coefficients of TbFe$_2$ and DyFe$_2$ have
not been calculated from first principles previously.
Experimental determination of these cubic terms is also expected
to be difficult due to the large magnetostrictive effect.
For example, based on M\"ossbauer measurements the authors of Ref.~\cite{Atzmony1973}
were unable to determine precise values of $[A_{lm}\langle r^l \rangle]$,
except that $A_{40}$ was of order 10~K/$a_0^4$.
Furthermore, the ratio $A_{60}/A_{40}$ was determined to
be -0.04$a_0^{-2}$.
As pointed out in Sec.~\ref{sec.BandA},  we cannot extract $A_{lm}$ from $\langle r^l \rangle$
so a direct comparison is not possible.
However, our value of $[A_{40}\langle r^4 \rangle]$ is also of order
10~K, and our calculated ratio $[A_{60}\langle r^6 \rangle]/[A_{40}\langle r^4 \rangle]$
is -0.08.
Therefore we can at least say that the sign of the ratio of the $l$ = 4 and 6 CF coefficients
is consistent with Ref.~\cite{Atzmony1973}, and also that the magnitudes are reasonable.

\subsubsection{Tb}
The final material we consider is elemental Tb.
Here, every atom in the system is replaced with Y, so arguably
this material presents the most difficult challenge to
the Y-analogue model, even though the nominal valence electron
configurations are the same.

Tb is hexagonal, and has the same nonzero CF coefficients as
NdCo$_5$.
Indeed, as seen in Table~\ref{tab.CFcoffs} the relative magnitudes
of the different CF coefficients of Tb follow those of NdCo$_5$ reasonably
closely, except for the crucial difference of
$[A_{66}\langle r^6 \rangle]$, which is positive for Tb.
Noting that Tb$^{3+}$ has negative Stevens $\alpha_J$ and $\gamma_J$
coefficients like Nd$^{3+}$, the calculated CF coefficients
lead to easy plane anisotropy, with a preferred direction
in the basal plane of $\phi = 0$, i.e.\ pointing between
the nearest in-plane Tb atoms [Fig.~\ref{fig.hex}(a)].
The calculated easy direction matches that observed experimentally~\cite{Radwanski1987}.

Unlike the other materials in our test set, there are no other magnetic
sublattices present in Tb.
Therefore, we can use the method described in Sec.~\ref{sec.KfromA} to 
calculate anisotropy constants at 0~K, obtaining values of -17, -12 and 5~MJm$^{-3}$ for $K_1$, $K_2$ and $K_3$,
and -0.2~MJm$^{-3}$ for $K_4$.
These numbers are smaller than the values reported from low temperature high-field measurements,
e.g.\ -60~MJm$^{-3}$ for $K_1$ only, and -2.4~MJ for $K_4$~\cite{Rhyne1967,Chikazumi1969}.
However it is important to remember that our calculations do not
include magnetostrictive effects, which are highly important in Tb.
For instance, the high-field magnetization curve clearly shows a hysteresis,
which has been attributed to a plastic deformation of the crystal~\cite{Chikazumi1969}.

\section{Conclusions and outlook}
\label{sec.conclusions}

Aside from the numerical results of our work contained
in Table~\ref{tab.CFcoffs}, we can summarize
our results qualitatively as follows:
\begin{itemize}
\item The signs of our calculated CF coefficients are consistent
with experimentally-observed/previously calculated magnetization directions
for all 10 materials (easy axis: SmCo$_5$, Sm$_2$Co$_{17}$, NdFe$_{12}$N, SmFe$_{12}$; 
easy plane: NdCo$_5$, NdFe$_{12}$, SmFe$_{12}$N, Tb; [111] axis: TbFe$_2$;
[001] axis, DyFe$_2$).
\item For systems with data available (NdCo$_5$ and Tb),
the signs of the $[A_{66} \langle r^6 \rangle]$ coefficients
are also consistent with the experimentally-observed basal plane anisotropy.
\item Our calculated CF coefficients are generally close (within 50~K/4~meV) of
previously reported open core calculations.
\item The trend across the Sm-Co series behaves intuitively, with
CF coefficients weakened for lower symmetry (SmCo$_5$
vs Sm$_2$Co$_{17}$).
\end{itemize}
Taken together, our results demonstrate that the 
Y-analogue model is a viable method of calculating CF coefficients
of RE-TM magnets.
Of course, it is not the only method, and we do not claim that it
is any more accurate than previously published 
work~\cite{Richter1998,Miyake2014,Delange2017,Richter1995,Novak1996,Steinbeck1996,
Harashima2015,Novak2013}.
In our opinion the strength of the Y-analogue model is in the simplicity
of the calculation of the CF potential $V({\bf r})$.
Whilst implemented in some
DFT codes like \texttt{OpenMX}~\cite{Ozaki2003}, the open-core approximation 
(which appears to give similar CF coefficients to the Y-analogue model), 
is not yet universally available.
By contrast, the PAW dataset for Y is distributed as standard 
in all of the popular DFT codes.
We also find the calculations involving Y to be numerically stable.
As such, we would expect a high degree of reproducibility of
the CF potential calculated for Y-analogues using different codes, similar to that
demonstrated e.g.\ for bulk moduli~\cite{Lejaeghere2016}.

As noted in Sec.~\ref{sec.theory}, the CF potential must be
supplemented by the radial 4$f$ charge density $n^0_{4f}(r)$  in order to calculate
the CF coefficient.
Unfortunately, calculating $n^0_{4f}(r)$ is not completely straightforward, requiring
a way of dealing with the 4$f$ electrons like the LSIC.
However, as we argued in the discussion surrounding Fig.~\ref{fig.felec}, 
the same  $n^0_{4f}(r)$ could be used to calculate the CF coefficients 
for a range of compounds containing a given lanthanide.
This further approximation would then mean only the CF potential
needed to be calculated for each compound,  enabling high throughput computational
screening of materials based on their calculated CF coefficients and scaling
up the work of e.g.\ Refs.~\cite{Korner2016,Tatetsu2018}.

From the point of view of assessing the predictive power of our calculations, 
the most interesting avenue for future exploration is to go beyond calculating
CF coefficients and instead target quantities that can be directly compared
to experiments, e.g.\ magnetization versus field curves, from which
anisotropy constants $K_i$ (\ref{eq.K}) are extracted.
Although we tentatively discussed anisotropy constants for Tb,
our ability to compare to experiment was limited due to us not
taking any magnetoelastic effects into account.
Including such effects will be crucial if we are to
properly account for the magnetostriction of the
heavy RE materials, including elemental Tb~\cite{Rhyne1967,Chikazumi1969}
and the Laves phase Tb$_{1-x}$Dy$_{x}$Fe$_2$ compounds~\cite{Clark1973}.

More generally, we did not attempt to determine $K_i$ for the RE-TM compounds
here since the CF coefficients do not include the contribution
to the anisotropy from the TM.
However, we have previously calculated 
the TM contribution for GdCo$_5$ (which has no crystal field effects) and obtained values
of $K_i$ in quantitative agreement with experiment, over a range of temperatures~\cite{Patrick2018}.
Therefore, by combining the method presented here with our previous work~\cite{Patrick2018} 
and finite temperature CF theory~\cite{Kuzmin2008},
we should have the necessary tools to calculate temperature-dependent 
anisotropy constants of RE-TM magnets entirely from first principles.

\section*{Acknowledgments}
The present work forms part of the PRETAMAG project,
funded by the UK Engineering and Physical Sciences Research
Council, Grant No. EP/M028941/1.
We thank J.\ J.\ Mortensen for guidance with the \texttt{GPAW} code.

\appendix
\section{Explicit expression for correction to the pseudo-density}
\label{app.density}

Here we give the explicit expression for the angular expansion 
of the correction to the pseudo-density.
We use standard \texttt{GPAW} notation~\cite{Enkovaara2010}.
The difference between the full and pseudized density $\Delta \rho({\bf r})$ is
the sum of the electronic contribution $\Delta n({\bf r})$
and a nuclear part:
\begin{equation}
\Delta \rho({\bf r}) = \Delta n({\bf r})
- Z \delta({\bf r}) - \sum_{lm} Q_{lm} g_{lm}({\bf r})
\end{equation}
Here, $Z$ is the nuclear charge (a positive number), $Q_{lm}$ are the
compensation charges used to ensure the multipole moments of  $\Delta \rho({\bf r})$
are zero, and $g_{lm}$ are functions which in \texttt{GPAW} are Gaussians
multiplied by real spherical harmonics.
The spin-resolved correction for the electronic density is
\begin{eqnarray}
\Delta n^\sigma({\bf r}) &=& \frac{n_c(r) - \tilde{n}_c(r)}{2}  + \nonumber \\
&&\sum_{i_1i_2}D_{\sigma i_1 i_2}[\phi_{i_1}({\bf r})\phi_{i_2}({\bf r}) -
\tilde{\phi}_{i_1}({\bf r})\tilde{\phi}_{i_2}({\bf r})].
\end{eqnarray}
Here, $n_c$ and $\tilde{n}_c$ are the full and pseudo core densities.
$D$ is the atomic density matrix and $\phi$, $\tilde{\phi}$ are the partial
waves, which together allow the reconstruction of the wavefunction in the rapidly-varying
region close to the nucleus.
$i$ is a composite index standing for ($\nu, l,m$) where $\nu$ plays the role
of the principal quantum number of the partial waves.

$Z$, $g_{lm}$, $n_c$, $\tilde{n}_c$, $\phi$ and $\tilde{\phi}$ are properties of the
PAW dataset, while
$Q_{lm}$ and $D_{\sigma i_1 i_2}$ are determined in the self-consistent calculation.
Noting that the compensation charges and the partial waves have the forms
$g_{lm}({\bf r}) = g_{l}(r) Y^R_{lm}({\bf \hat{r}})$ and
 $\phi_{i_1}({\bf r}) = \phi_{\nu l}(r) Y^R_{lm}({\bf \hat{r}})$ respectively,
with the $R$ denoting real spherical harmonics, the angular expansions of
$\Delta \rho$ and $\Delta n^\sigma$ are readily obtained.
The only slight complication is the use of real spherical
harmonics in \texttt{GPAW}, which means that one must take some care
with integrals like $\int d{\bf \hat{r}} Y^*_{lm}({\bf \hat{r}})Y^R_{l'm'}({\bf \hat{r}})$
and $\int d{\bf \hat{r}} Y^*_{lm}({\bf \hat{r}})Y^R_{l_1m_1}({\bf \hat{r}})Y^R_{l_2m_2}({\bf \hat{r}})$.
\section*{References}

\providecommand{\newblock}{}

\end{document}